\documentclass[10pt,preprint,showpacs,preprintnumbers,amsmath,amssymb]{revtex4}

\usepackage{graphicx}% Include figure files
\usepackage{dcolumn}% Align table columns on decimal point
\usepackage{bm}% bold math
\usepackage{slashed}
\begin{document}

%\date{\today}

\title{Divergent energy strings in $AdS_5\times S^5$ with three angular momenta}

\author{Sergio Giardino}
 \email{jardino@fma.if.usp.br}
\affiliation{Instituto de F\'{i}sica, Universidade de S\~{a}o Paulo, Brazil}
%,\\ CP 66318, 05315-970 S\~{a}o Paulo, SP, Brazil.}

\begin{abstract}
\noindent In this paper, novel solutions for strings with three angular
momenta in $AdS_5 \times S^5$ geometry are presented; the divergent
energy limit and the corresponding conserved charges, as well as dispersion
relation are also determined. Interpretations 
of these configurations as either a giant magnon (GM) or a spiky
string (SS) are
discussed. 
\end{abstract}

\maketitle

\section{Introduction}

Divergent energy solutions have an important role in classical
string theory. Such strings can be interpreted in
terms of the gauge theory, but even in cases where an exact interpretation
is lacking, there are enough characteristic features in these solutions to
characterize them in a class of their own. The most well-understood cases concern solutions known as giant magnons
\cite{Hofman:2006xt}, and when one string of this kind has one angular momentum in the $\mathbb{R}\times S^2$
subspace of $AdS_5\times S^5$, it obeys the dispersion relation
\begin{equation} 
E-J =2T \left|\sin\frac{p}{2}\right|, \label{E-J}
\end{equation}
\noindent where {\it E, J } and {\it T} are the energy, the
angular momentum, and the tension of the string, respectively. {\it p} is
identified with the angle between the extremes of the giant magnon
and $E,\,J\to\infty$. In another case, a solution known as
spiky string \cite{Kruczenski:2006pk,Kruczenski:2004wg} also
describes a divergent energy string with one angular momentum and it
too obeys the dispersion relation 
\begin{equation}
E-T\Delta\phi=2T\left(\frac{\pi}{2}-\theta\right), \label{SS_zero}
\end{equation}
\noindent where {\it T} is the string tension,  $\Delta\phi\to \infty$
is the deficit angle and $\theta$ is the coordinate where the string
peaks.

Variations in dispersion relations can be found in strings
rotating in various dimensions and in different geometries. For
example, giant magnons are found in $AdS_5\times S^5$ with two
\cite{Kruczenski:2006pk,Bobev:2006fg,Dimov:2007ey,Vicedo:2007rp,Chen:2006gea,Lee:2008yq}
and three \cite{Kruczenski:2006pk,Dimov:2007ey} angular momenta. In
the Lunin-Maldacena background, GM's are found with one \cite{Bykov:2008bj,Bobev:2007bm} and two 
\cite{Bykov:2008bj,Chu:2006ae,Bobev:2007bm} angular momenta; and in 
$AdS_4\times \mathbb{CP}^3$, with one
\cite{Lee:2008yq,Suzuki:2009sc,Abbott:2008qd,Gaiotto:2008cg}, two
\cite{Lee:2008ui,Lee:2008yq,Ryang:2008rc,Kalousios:2010ne,Abbott:2008qd,Grignani:2008is,Berenstein:2008dc}
and three \cite{Giardino:2011dz} angular momenta.

In the case of spiky strings different examples have already been found.
In $AdS_5\times S^5$ geometry, spiky strings have been found with one \cite{Ishizeki:2007we}, two
\cite{Bobev:2007bm} and three \cite{Dimov:2007ey} angular momenta; and
in $AdS_4\times \mathbb{CP}^3$ with one \cite{Lee:2008ui},
two \cite{Ryang:2008rc} and three \cite{Giardino:2011dz} angular momenta.

The more angular momenta are added to the string, the more the
dispersion relation changes. If we have two angular momenta, the
dispersion relation \cite{Dorey:2006dq, Minahan:2006bd}
\begin{equation}
E-J_1=\sqrt{J_2^2+\frac{\lambda}{\pi^2}\sin^2\frac{p}{2}}
\end{equation}
describes a giant magnon string in $AdS_5\times S^5$ where $E$ and
$J_1$ are divergent and $J_2$ is finite. If we allow more divergent
quantities, the complexity increases, and we find the following dispersion
relation \cite{Kluson:2007qu}
\begin{equation}
\sqrt{E^2-J_\phi^2}-J_\psi=T\sin\Delta\psi\label{DR1}
\end{equation}
\noindent found in giant magnons in $AdS_4\times \mathbb{CP}^3$
\cite{Ryang:2008rc,Giardino:2011dz}, and in the case of many divergent
angles the spiky string-like dispersion relation can be even more complex \cite{Giardino:2011dz}.
The interpretation of these solutions in the case of gauge theory is
unknown, but if the gauge/gravity correspondence is correct, these
solutions also have a dual description. 

In a previous study \cite{Giardino:2011dz}, we addressed several
multi-divergent solutions for strings in the $AdS_4\times
\mathbb{CP}^3$ background with three angular momenta. These solutions
are important in order to answer the question of the existence of giant magnons
and spiky strings with three angular momenta in the aforementioned background. In
the case of $AdS_5\times S^5$, the references cited above include
giant magnons and spiky strings with various angular momenta, but not
with various divergent angular momenta and deficit angles, and the
purpose of this
article is to fill in this void in current literature. In some sense
we generalize the former results.

The contents of this paper are organized as follows: in the second section,  multi-divergent solutions are constructed in their
full generality, while  in the third and fourth sections, we take the
particular cases where divergent energy is found and interpret them in terms of
giant magnons and spiky strings. The fourth section presents our conclusions.

\section{The general solution}
We start with the complete $AdS_5\times S^5$ metric
\begin{eqnarray}
ds^2 & = &R^2\left(-\cosh^2\rho\, dt^2 + d\rho^2+\sinh^2\rho \,d\Omega^2_3\right)+\nonumber \\  
& + & R^2\left(d\theta^2+\sin^2\theta\,d\psi^2 +\sin^2\theta\cos^2\psi\,
  d\phi_1^2 +\cos^2\theta\,d\phi_2^2 + \sin^2\theta\sin^2\psi\, d\phi_3^2 \right).\label{mtr:inic}
\end{eqnarray}
\noindent The first term of expression (\ref{mtr:inic}), in brackets,
corresponds to the metric of a five dimensional anti-de 
Sitter space whose elements are time, $t$, a radial coordinate,
$\rho$, and a three dimensional sphere. The second term in brackets is
a metric of a five dimensional sphere, whose coordinates
are parameterized as $\theta,\,\psi\in[0,\,\pi/2]$ and
$\phi_{i=\{1,\,2,\,3\}}\in[0,\,2\pi]$. We are seeking solutions at the
center of the anti-de Sitter space, which means that $\rho=0$ and the
string is confined in a $\mathbb{R}\times S^5$ subspace. The motion of
the string
 we are interested in has four degrees of freedom, as the coordinate $\theta$ is
kept constant. Of course, if we chose $\psi$ as the constant, the
results would be exactly the same. Thus, we chose the following ansatz for the varying coordinates
\begin{equation}
t=\kappa\tau, \qquad \theta\in(0,\,\pi/2), \qquad \psi=\psi(y), \qquad \phi_{i=\{1,\,2,\,3\}}=\omega_i \tau + f_i(y)
\label{ansatz:1}
\end{equation}
\noindent where $y=\alpha\sigma + \beta \tau$, and $\alpha$, $\beta$,
$\kappa$, and $\omega_i$ are constants, and $\sigma$ and
$\tau$ parameterize the dynamics of the world sheet of the string. The
dynamics of a string with tension $T=\frac{\sqrt{\lambda}}{2\pi}$
can be described by Polyakov action 
\begin{equation}
S=\frac{T}{2}\int d\sigma\,d\tau\,\gamma^{a\,b} g_{\mu\,\nu}\,\partial_a X^{\mu}\,\partial_b X^{\nu},
\end{equation}
\noindent and the Virasoro constraints
\begin{equation}
g_{\mu\nu}\partial_{\tau} X^{\mu}\partial_{\sigma}X^{\nu}=0\hspace{5mm} \textnormal{and} \hspace{5mm}
g_{\mu\nu}\left(\partial_{\tau}X^{\mu}\partial_{\tau}X^{\nu} +  \partial_{\sigma}X^{\mu}\partial_{\sigma}X^{\nu} \right)=0, \label{Virasoro}
\end{equation}
\noindent where $\gamma^{a\,b}=(-1,-,1)$ is the world sheet metric.

\noindent Using (\ref{ansatz:1}) and the equations of motion for $\phi_i$, we
obtain
\begin{equation}
\partial_y f_i=\frac{1}{\alpha^2-\beta^2}\left(\frac{A_i}{g_i} +\beta\omega_i \right)\label{int:f}
\end{equation}
\noindent where $A_i$ is the integration constant, and $g_i$ the metric
tensor component corresponding to $\phi_i$. Using  (\ref{int:f}) and the Virasoro
constraints (\ref{Virasoro}) we get
\begin{equation}
\psi_y^2+w_1^2\cos^2\psi+w_2^2+w_3^2\sin^2\psi+\frac{a_1^2}{\cos^2\psi}+a_2^2+\frac{a_3^2}{\sin^2\psi}-k^2=0\label{psi_y}
\end{equation}
\noindent such that $\psi_y=\partial_y\psi$ and the equation
constants come from a redefinition of the older ones, namely\begin{eqnarray}
&&a_1=\frac{1}{\alpha^2-\beta^2}\frac{A_1}{\sin^2\theta},\qquad
a_2=\frac{1}{\alpha^2-\beta^2}\frac{A_2}{\cos^2\theta}, \qquad
a_3=\frac{1}{\alpha^2-\beta^2}\frac{A_3}{\sin^2\theta},\nonumber\\
&& \nonumber \\
&& w_1=\sin^2\theta\frac{\alpha\,\omega_1}{\alpha^2-\beta^2},\qquad
w_2=\cos^2\theta\frac{\alpha\,\omega_2}{\alpha^2-\beta^2}\qquad
w_3=\sin^2\theta\frac{\alpha\,\omega_3}{\alpha^2-\beta^2} \nonumber \\
&& \nonumber \\
&& \mbox{and} \qquad  k^2=\frac{\alpha^2+\beta^2}{\left(\alpha^2-\beta^2\right)^2}\,\kappa^2.\nonumber
\end{eqnarray}
\noindent With the change of variables $\mathcal{X}=\cos 2\psi$
so that (\ref{psi_y}) becomes
\begin{eqnarray}
\frac{1}{16}\mathcal{X}_y^2&=&\frac{w_1^2-w_3^2}{8}\,\mathcal{X}^3+\left(\frac{w_1^2+w_3^2}{8}-\frac{m}{4}
\right)\mathcal{X}^2+\nonumber\\
&+&\left(\frac{w_3^2-w_1^2}{8}+\frac{a_1^2-a_3^2}{2}\right)\mathcal{X}-\frac{w_1^2+w_3^2}{8}-\frac{a_1^2+a_3^2}{2}+\frac{m}{4},\label{Xy}
\end{eqnarray}

\noindent and $m=k^2-w_2^2-a_2^2$. Equation (\ref{Xy}) shows that
 the polynomial degree drops from three to two, if
$w_1^2=w_3^2$. Thus, we study the two different cases separately.

\subsection{$w_1^2\neq w_3^2$}
To obtain divergent energy, we
choose
 \begin{equation}
a^2_{i=1,\,3}=\frac{1}{4}\left(m-w_i^2\right).
\end{equation}
\noindent The energy is real if $w_3^2>w_1^2$, and so finally we get
\begin{equation}
dy=\frac{1}{\sqrt{2\,(w_3^2-w_1^2)}}\frac{d\mathcal{X}}{\mathcal{X}\sqrt{\mathcal{X}_0-\mathcal{X}}},\label{dy}
\end{equation}
\noindent where $\mathcal{X}_0\in (0,\,1)$ is given by
\begin{equation}
\mathcal{X}_0=\frac{w_1^2+w_3^2-2m}{w_3^2-w_1^2}.
\end{equation}

\noindent Expression (\ref{dy}) can be used to calculate the conserved
charges
\begin{equation}
E=T\int\limits_{0}^{2\pi}d\sigma\,\dot{t}, \qquad \mbox{and}\qquad
J_i=T\int\limits_{0}^{2\pi}d\sigma\,g_i\,\dot{\phi_i}\label{J+E}.
\end{equation}
\noindent Using $d\sigma=\alpha\,dy$ and
$\mathcal{X}\in[0,\mathcal{X}_0]$, the energy is divergent and the
momenta and the deficit angles are 
\begin{eqnarray}
&&\mathcal{J}_{i=1,\,3}=\frac{J_i}{\sin^2\theta}=\left(\beta\,a_i+\frac{\alpha \,
  w_i}{2}\right)\frac{E}{\kappa}\pm \frac{T\,w_i}{\sqrt{2\,(w_3^2-w_1^2)}}\,I_0 \nonumber\\
&& \Delta\phi_{i=1,\,3}=\left(\alpha\,a_i+\beta
  w_i\right)\frac{E}{\kappa\,
  T}+\frac{2\,a_i}{\sqrt{2\,(w_3^2-w_1^2)}}\,I_i \label{SOL}\\
&&\mathcal{J}_2=\frac{J_2}{\cos^2\theta}=\left(\beta\,a_2+\alpha \,
  w_2\right)\frac{E}{\kappa}\qquad\mbox{and}\qquad\Delta\phi_2=\left(\alpha\,a_2+\beta
  w_2\right)\frac{E}{\kappa\,T},\nonumber
\end{eqnarray}
\noindent where 
\begin{eqnarray}
&&I_0=\intop_0^{\mathcal{X}_0}\frac{d\mathcal{X}}{\sqrt{\mathcal{X}_0-\mathcal{X}}}=2\sqrt{\mathcal{X}_0}\nonumber\\
&&I_1=\intop_0^{\mathcal{X}_0}\frac{1}{\mathcal{X}-
  1}\frac{d\mathcal{X}}{\sqrt{\mathcal{X}_0-\mathcal{X}}}=-\frac{2}{\sqrt{1-\mathcal{X}_0}}\arctan\sqrt{\frac{\mathcal{X}_0}{1-\mathcal{X}_0}}\label{III}\\
&&I_3=\intop_0^{\mathcal{X}_0}\frac{1}{\mathcal{X}+
  1}\frac{d\mathcal{X}}{\sqrt{\mathcal{X}_0-\mathcal{X}}}=\frac{2}{\sqrt{1+\mathcal{X}_0}}\,\mbox{arctanh}\sqrt{\frac{\mathcal{X}_0}{1+\mathcal{X}_0}}\nonumber
\end{eqnarray}
\noindent Expressions (\ref{SOL}) and (\ref{III}) summarize all the information about the
conserved charges of the system. We point out that if
$\mathcal{X}_0\geq 1$, all the charges are divergent without a finite term, so
this case does not generate either giant magnons or spiky strings.
\subsection{$w_1^2 = w_3^2$}
In this case, we use $\cos2\psi=\mathcal{Y}$, $w_1^2 = w_3^2=w$ and so (\ref{Xy}) changes to
\begin{equation}
\frac{1}{8}\mathcal{Y}_y^2=-\frac{n}{2}\mathcal{Y}^2+\left(a_1^2-a_3^2\right)\,\mathcal{Y}+\frac{n}{2}-a_1^2-a_3^2,\label{Yy}
\end{equation}
\noindent so that $n=m-w^2$. Divergent energy is obtained as long as
$n=n_\pm=\left(a_1\pm a_3\right)^2$, and in this case (\ref{Yy}) spans
\begin{equation}
dy=\frac{1}{2\sqrt{n}}\,\frac{d\mathcal{Y}}{\mathcal{Y}_0-\mathcal{Y}},
\end{equation}
\noindent so that $\mathcal{Y}_0\in(0,\,1]$, because if
$\mathcal{Y}_0>1$ there is no divergence in the integral. Thus, either
\begin{equation}
\mathcal{Y}_0=\frac{a_1-a_3}{a_1+a_3} \qquad \mbox{if}\qquad
n=n_+,\qquad\mbox{otherwise} \qquad\mathcal{Y}_0=\frac{a_1+a_3}{a_1-a_3}.
\end{equation}
The conserved charges to this string are

\begin{eqnarray}
&&\mathcal{J}_{i=1,3}=\left(\beta
  a_i+\alpha\,
  w\,\frac{1\pm\mathcal{Y}_0}{2}\right)\frac{E}{\kappa}\pm
T\,w\,\mathcal{Y}_0\nonumber\\
&&\Delta\phi_i=\left(\beta w+\frac{2\,\alpha\,a_i }{1\pm
    \mathcal{Y}_0}\right)\frac{E}{\kappa\,T}+\frac{2\,a_i}{1\mp\mathcal{Y}_0}\,\ln(1\pm \mathcal{Y}_0).
\end{eqnarray}
\noindent $\mathcal{J}_2$, and $\Delta\phi_2$ have equal expressions
as (\ref{SOL}).

\subsection{Common features }
We classify a solution according to the dispersion relations it
generates, and we claim that both cases have giant magnons and spiky
strings described by similar dispersion relations. What is meant by similar
is that the relation among the divergent quantities is totally equal,
with only the finite term being different, as it is generated by the
particular constraints among the constants of the problem. We justify
our claim  by establishing the following equivalence relations:
\begin{eqnarray}
\left(\mathcal{J}_1+\mathcal{J}_3\right)_{w_1^2\neq w_3^2} &\sim&
\left(\mathcal{J}_1-\mathcal{J}_3\right)_{w_1^2= w_3^2}\label{Req1}\\
\left(\mathcal{J}_2\right)_{w_1^2\neq w_3^2} &\sim&\left(\mathcal{J}_1+\mathcal{J}_2+\mathcal{J}_3\right)_{w_1^2= w_3^2}.\label{Req2}
\end{eqnarray}
\noindent Expression (\ref{Req1}) relates quantities formed by a
finite and a divergent term, and (\ref{Req2}) relates two 
intrinsically divergent quantities. Analogous considerations can
be carried out with regards to
the deficit angles. So, the relations that can be built among these
quantities are the same in both the equivalence classes. Of course, the
finite term will  a have different expression, but with the
same physical meaning. Also, the cases are physically equivalent, and
in the following sections, we study different dispersion relations
by considering  only the most general situation, where $w_1^2\neq w_3^2$.
\section{Giant magnon solutions}

Here we need finite deficit angles and divergent energy and
momenta, and we obtain these by choosing values for the constants in
(\ref{SOL}). The most obvious possibility, $\alpha a_i=-\beta w_i$,
forces $w_1^2=w_3^2=0$, which is not acceptable. So, we pick a
more general condition
\begin{equation}
\alpha (a_1+a_3)=-\beta (w_1+w_3),\qquad \alpha a_2=-\beta w_2
\qquad\mbox{and}\qquad \alpha\, a_{1,\,3}\neq-\beta w_{1\,3},\label{Cmg}
\end{equation}
\noindent so that, for $\mathcal{J}=\mathcal{J}_1+\mathcal{J}_3$ and $\Delta\phi=\Delta\phi_1+\Delta\phi_3$
\begin{eqnarray}
&&\mathcal{J}=\alpha\left(\omega_1+\omega_3\right)\left(\frac{1}{2}-\frac{\beta^2}{\alpha^2}\right)\frac{E}{\kappa}+T\left(\omega_1-\omega_3\right)\,I_0,\nonumber\\
&&\mathcal{J}_2=\alpha\omega_2\left(1-\frac{\beta^2}{\alpha^2}\right)\frac{E}{\kappa}\label{MG}\\
&&\Delta\phi=2\left(a_1\,I_1+a_3\,I_3\right)\qquad \Delta\phi_2=0.\nonumber
\end{eqnarray}
A corresponding set of momenta and divergent angles has already been found in
$AdS_4\times\mathbb{CP}^3$ \cite{Ryang:2008rc,Giardino:2011dz} with a particular dispersion
relation akin to (\ref{DR1}). Hence (\ref{MG}) describes a solution that is a
giant magnon of this kind. The only thing to do is to choose the correct
constraints among the parameters of the problem. The compatibility
between the difference of the Virasoro constraints and a relation that
comes from the dispersion relation results in a constraint which permits us to
determine $a_1$ as

\begin{equation}
a_1=\frac{w_1+w_3}{w_3-w_1}\frac{\beta}{\alpha}\left[\left(\frac{\frac{1}{2}-\frac{\beta^2}{\alpha^2}}{1-\frac{\beta^2}{\alpha^2}}\right)^2
  (w_1+w_3)-w_3\right].\label{A1ss}
\end{equation}
\noindent In addition, as $\mathcal{X}_0\in(0,\,1)$, we have
\begin{equation}
w_2^2\left(1-\frac{\beta^2}{\alpha^2}\right)+w_1^2<k^2.
\end{equation}
\noindent The deficit angle has a constraint
which allows it to be related to the finite term in the
dispersion relation as follows:
\begin{equation}
\sin\Delta\phi=(w_1-w_3)\,I_0.\label{vinc_mg}
\end{equation}
\noindent Relations (\ref{A1ss}), (\ref{Cmg}), and (\ref{vinc_mg}) allow
us to eliminate the integration constants and so define a complex
constraint among the parameters of the problem, which we not include here
because it does not contribute anything with regards to the physics of the problem. 
All the conditions above allow us to write the dispersion relation of
the giant magnon as
\begin{equation}
\sqrt{E^2-\mathcal{J}_2^2}-\mathcal{J}=T\sin\Delta\phi.\label{DR2}
\end{equation}

\noindent The above solution (\ref{DR2}), which has a known giant
magnon dispersion relation
\cite{Ryang:2008rc,Giardino:2011dz,Kluson:2007qu}, also has the presumed
feature that the limit
$\mathcal{J}_2\to0$ recovers the usual giant magnon dispersion
relation. Compared to the former solution of a giant magnon with
three angular momenta \cite{Kruczenski:2006pk}, we stress that all
three angular momenta of (\ref{DR2}) are divergent, while the former
solution has one divergent angular momentum and two finite angular
momenta. On the other hand, both have the same
limit when two angular momenta are taken to zero and only one
divergent angular momentum remains. Thus, the above result and
\cite{Kruczenski:2006pk} are different generalizations of the usual giant
magnon with one divergent angular momentum. Neither of these
generalizations is understood on the gauge side.

We can also obtain a curious giant magnon solution simply by
choosing different constants in the $\phi_2$ direction, such that
$\mathcal{J}_2=0$, and $\Delta\phi_2$ is divergent. The dispersion
relation in this case is given by
\begin{equation}
\sqrt{E^2-(T\,\Delta\phi_2)^2}-\mathcal{J}=T\sin\Delta\phi.\label{DR4}
\end{equation}
A corresponding case, where a deficit angle and a momenta play interchanged
roles has already been discovered in a spiky string by \cite{Ryang:2008rc}. The
effect of simply changing the momentum to the deficit angle in the
same direction does not have an interpretation, but (\ref{DR4}) certainly
describes a GM because the deficit angle and the subtracting momentum
belong to the same coordinate, and the limit $\Delta\phi_2\to 0$
recovers the GM dispersion relation.

\section{Spiky string solutions}

This case is built from divergent deficit angles and finite
momenta, so that the analysis  of the giant magnon case is valid here,
only changing $J$ by $T\,\Delta\phi$ in the appropriate places. As for
the giant magnon case, we cannot make the angular momenta finite in an
independent manner, because this implies $w^2_{i=1,3}=0$. Thus we impose 
\begin{equation}
2\beta (a_1+a_3)=-\alpha (w_1+w_3),\qquad \beta a_2=-\alpha w_2
\qquad\mbox{so that}\qquad \alpha\, a_{1,\,3}\neq-\beta w_{1\,3},
\end{equation}
\noindent to obtain, with the very notation of the
preceding section, that
\begin{eqnarray}
&&\mathcal{J}=\sqrt{2\,\frac{w_3+w_1}{w_3-w_1}}\,I_0, \qquad
\mathcal{J}_2=0,\\
&&\Delta\phi=\left(1-\frac{\alpha^2}{2\beta^2}\right)\beta\,\left(w_1+w_3\right)\,\frac{E}{\kappa\,T}+\sqrt{\frac{2}{w_3^2-w_1^2}}\left(a_1\,I_1+a_3\,I_3\right)\\
&&\Delta\phi_2=\left(1-\frac{\alpha^2}{\beta^2}\right)\beta\,w_2\,\frac{E}{\kappa\,T}.
\end{eqnarray}
\noindent As $\mathcal{X}_0\in(0,\,1)$, we also write that
\begin{equation}
w_2^2\left(1+\frac{\alpha^2}{\beta^2}\right)+w_1^2<k^2.
\end{equation}
\noindent And if we choose
\begin{equation}
a_1=\frac{w_1+w_3}{w_3-w_1}\frac{\alpha}{\beta}\left[\left(\frac{1-\frac{\alpha^2}{2\beta^2}}{1-\frac{\alpha^2}{\beta^2}}\right)^2
  (w_1+w_3)-\frac{w_3}{2}\right].\label{A1}
\end{equation}
\noindent we write a dispersion relation to a spiky string
\begin{equation}
\sqrt{E^2-\left(T\,\Delta\phi_2\right)^2}-T\,\Delta\phi=2T\left(\frac{\pi}{2}-\psi_0\right),\label{SS1}
\end{equation}
\noindent so that
\begin{eqnarray}
&&\psi_0=\frac{\pi}{2}-\sqrt{\frac{2}{w_3^2-w_1^2}}(a_1\,I_1+a_3\,I_3)\nonumber\\
&&\cos2\psi_0=\mathcal{X}_0.\label{vinc:SS}
\end{eqnarray}
This is a very interesting result: it is totally analogous to the
giant magnon case, and it is a novel dispersion relation. A similar case
can be obtained from \cite{Giardino:2011dz} by imposing suitable
constraints, although this was not done by the authors. As in the
preceding case, the limit $\Delta\phi_2\to 0$ generates the usual
spiky string dispersion relation, and the unusual form of (\ref{SS1})
comes from the fact that there are more geometric degrees of freedom to the
motion of the string, and this greater geometric freedom requires a
less parametric freedom, thus complicating the constraints
(\ref{vinc:SS}) involving the constants of the model. When comparing 
(\ref{SS1}), which has two finite angular momenta and three divergent deficit
angles, to the spiky string found in \cite{Ryang:2008rc}, which has one
divergent angular momentum and one divergent deficit angle, we see
that both the results have a usual spiky string as a limit when
only one deficit angle remains as a divergent
quantity. Thus both cases have identical limits, even
though the strings are in different geometries. Accordingly, we infer
that these solutions probably have the same physical interpretation.

As for the giant magnon, we can have a different dispersion relation to
the spiky string. When we choose appropriate values for the constants, we get
\begin{equation}
\sqrt{E^2-\mathcal{J}_2^2}-T\,\Delta\phi=2T\left(\frac{\pi}{2}-\psi\right).\label{SS2}
\end{equation}
\noindent A similar situation appears is evident in $AdS_4\times\mathbb{CP}^3$ geometry, as
  studied by \cite{Ryang:2008rc}. As previously stated with regards to the GM case, we do not
  have an interpretation for the interchange of  momenta and the
  deficit angle, but it is certainly a spiky string solution because
  it behaves thus in the $\mathcal{J}_2\to 0$ limit.

To round off this section, we will mention that a dispersion relation
where both a giant magnon and a spiky string appears coupled, as 
the case in \cite{Giardino:2011dz}, does not seem to
be possible in any choice of constants for the strings described in
this article. The reason for this is that in order to construct a dispersion
relation akin to that found in the reference would require 
quantities in $\phi_2$ which are not intrinsically divergent, and
$\mathcal{J}_2$ and $\Delta\phi_2$ in (\ref{SOL}) do not have a finite
term to fulfill this feature

\section{Concluding remarks}

In this article we have presented new giant magnon and spiky string
solutions. These solutions are different from the usual giant magnon and
spiky string cases because they have divergences in various
dimensions. Naturally, this feature changes the dispersion relation of
the conserved charges, but the interpretation of these objects in the
gauge side of the duality remains, in principle, the same. A point
which supports this hypothesis is the satisfactory behavior of the
solutions, which recover lower dimensional GM and SS in the
appropriate limits. This
assumption also relies on the AdS/CFT conjecture, and although it has not yet
been proven, the dispersion relations found here must be found in the
dual gauge according to the correspondence hypothesis. We expect 
future studies in the gauge side of the correspondence to confirm this
prediction.
\linebreak
\paragraph*{\bf Acknowledgement:} The author is grateful for
the support offered by the Departamento de F\'{i}sica Matem\'{a}tica.
 
%%%%%%%%%%%%%%%%%%%%%%%%
%
%
%  BIBLIOGRAPHY
%
%
\bibliographystyle{unsrt} 
\bibliography{bib_mg_ads}
%%%%%%%%%%%%%%%%%%
%
%
%    END
%
%
\end{document}